# Survival of itinerant excitations and quantum spin state transitions in YbMgGaO$_4$ with chemical disorder

X. Rao[1,10], G. Hussain[1,10], Q. Huang[2,10], W. J. Chu[1], N. Li[1], X. Zhao[3], Z. Dun[2], E. S. Choi[4], T. Asaba[5], L. Chen[5], L. Li[5], X. Y. Yue[6], N. N. Wang[7], J.-G. Cheng[7], Y. H. Gao[8], Y. Shen[8], J. Zhao[8], G. Chen[9]★, H. D. Zhou[2]★, and X. F. Sun[1,6]★

[1]Hefei National Laboratory for Physical Sciences at Microscale, Department of Physics, and Key Laboratory of Strongly-Coupled Quantum Matter Physics (CAS), University of Science and Technology of China, Hefei, Anhui 230026, People's Republic of China

[2]Department of Physics and Astronomy, University of Tennessee, Knoxville, Tennessee 37996-1200, USA

[3]School of Physical Sciences, University of Science and Technology of China, Hefei, Anhui 230026, People's Republic of China

[4]National High Magnetic Field Laboratory, Florida State University, Tallahassee, FL 32310-3706, USA

[5]Department of Physics, University of Michigan, Ann Arbor, Michigan 48109, USA

[6]Institute of Physical Science and Information Technology, Anhui University, Hefei, Anhui 230601, People's Republic of China

[7]Beijing National Laboratory for Condensed Matter Physics and Institute of Physics, Chinese Academy of Sciences, Beijing 100190, People's Republic of China

[8]State Key Laboratory of Surface Physics and Department of Physics, Fudan University, Shanghai 200433, People's Republic of China

[9]Department of Physics and HKU-UCAS Joint Institute for Theoretical and Computational Physics at Hong Kong, The University of Hong Kong, Hong Kong, China

[10]These authors contributed equally: X. Rao, G. Hussain, Q. Huang

★email: gangchen.physics@gmail.com; hzhou10@utk.edu; xfsun@ustc.edu.cn

**Abstract**

**A recent focus of quantum spin liquid (QSL) studies is how disorder/randomness in a QSL candidate affects its true magnetic ground state. The ultimate question is whether the QSL**

survives disorder or the disorder leads to a "spin-liquid-like" state, such as the proposed random-singlet (RS) state. Since disorder is a standard feature of most QSL candidates, this question represents a major challenge for QSL candidates. YbMgGaO$_4$, a triangular lattice antiferromagnet with effective spin-1/2 Yb$^{3+}$ ions, is an ideal system to address this question, since it shows no long-range magnetic ordering with Mg/Ga site disorder. Despite the intensive study, it remains unresolved as to whether YbMgGaO$_4$ is a QSL or in the RS state. Here, through ultralow-temperature thermal conductivity and magnetic torque measurements, plus specific heat and DC magnetization data, we observed a residual $\kappa_0/T$ term and series of quantum spin state transitions in the zero temperature limit for YbMgGaO$_4$. These observations strongly suggest that a QSL state with itinerant excitations and quantum spin fluctuations survives disorder in YbMgGaO$_4$.

**Introduction**

A quantum spin liquid (QSL) is an exotic quantum state in which spins are highly entangled and remain disordered down to zero temperature limit[1-4]. It has attracted intensive interests because of its potential relevance to high temperature superconductivity and quantum information applications. The most fascinating feature of certain gapless QSL is that its fermionic-like spin excitations, or spinons, behave like mobile charge carriers in a paramagnetic metal with a Fermi surface[1-4]. Since the first reported gapless QSL candidate $EtMe_3Sb[Pd(dmit)_2]_2$[5], a spin-1/2 triangular lattice antiferromagnet (TAF) in 2010, the search for other candidates remains as a hot topic during the last decade. One commonly used experimental probe for QSL is a broad continuous magnetic excitation, or continuum mode, in the inelastic neutron scattering (INS) spectrum[6-8].

Meanwhile, the community recently started to pay attention to the chemical disorder effects on quantum magnetism. For example, some recent studies proposed that the randomness in a quantum magnet can induce spin-singlet dimers of varying strengths with a spatially random manner, which can account for the continuum mode due to its widely distributed binding energy. Indeed, disorder is unavoidable in most of the studied gapless QSL candidates. For instance, the kagome lattice herbertsmithite $ZnCu_3(OH)_6Cl_2$ has $Zn^{2+}/Cu^{2+}$ site mixture[9] (whether there is a small gap in this QSL candidate is still under debate); the $LiZn_2Mo_3O_8$ with breathing kagome lattice has $Li^+/Zn^{2+}$ site mixture[10]; the $Ca_{10}Cr_7O_{28}$ with bilayer kagome lattice has disorder among the two different $Cr^{3+}$ positions[11,12]; and the $H_3LiIr_2O_6$ with honeycomb lattice has mobile Hydrogen ions[13]. Therefore, this so-called random-singlet (RS) or valence bond glass (VBG) state[14-21] seriously prompts re-consideration of the intrinsic magnetic ground state of them: whether they are true gapless QSL or just spin liquid like?

$YbMgGaO_4$ (YMGO), a TAF with the effective spin-1/2 $Yb^{3+}$ ions[22,23], is in the center of this controversy. On one hand, the observed continuum mode[7,8,24,25], the $C_p \sim T^{0.7}$ behavior for the specific heat[22], the temperature-independent plateau for the Muon spin relaxation (MuSR) rate[26], and the saturated DC susceptibility below 0.1 ~ 0.2 K[28] all suggest a gapless QSL state. Its origin has been interpreted as a U(1) QSL state with spinon Fermi surface[8,25,28,29], or a resonant valence bond (RVB) like state[24], or a $J_1$-$J_2$ exchange interactions resulted QSL[7,30]. On the other hand, no residual $\kappa_0/T$ term on the thermal conductivity has been observed[31]. It needs to be emphasized that since the itinerant spin excitations, spinons, carry entropy[5,32-34], a non-zero residual linear $\kappa_0/T$ term at ultralow temperature for thermal conductivity should exist. It is a well-recognized signature for spinon. The absence of the $\kappa_0/T$ term in YMGO makes it difficult to reconcile with the spinon physics. The frequency dependent AC susceptibility peak[35] further suggests that the disordered occupancy of the inter-triangular layers by $Mg^{2+}$ and $Ga^{3+}$ ions[23,36] lead to a glassy ground state. By including this disorder, other types of ground states have been proposed, such as the mimicry of a spin liquid[37,38], RS led spin-liquid-like behavior[16,39], and randomness induced spin-liquid-like phase in $J_1$-$J_2$ model[40].

In this paper, with the motivation to settle down this dispute, which is certainly significant for QSL studies, we performed thermal conductivity, specific heat, magnetic torque, DC magnetization, and AC susceptibility measurements on high-quality YMGO single crystals at ultralow temperatures with two crystallographic axes and detailed field scans to study its intrinsic magnetic ground state. We observed a residual $\kappa_0/T$ term and series of quantum spin state transitions in the zero temperature limit, which strongly suggest that a QSL state with itinerant excitations and quantum spin fluctuations survives disorder in YMGO.

## Results and discussions

**Thermal conductivity.** Figure 1a shows the ultralow-temperature thermal conductivity ($\kappa$) of YMGO measured along either the *a* or *c* axis, plotted with $\kappa/T$ vs *T*. Apparently, the $\kappa_a$ displays a $T^2$ behavior in a rather broad temperature range (200 – 600 mK) with a slope change at $T < 200$ mK. As shown in Fig. 1a, the $\kappa_c$ also exhibits a $T^2$ behavior at $T < 600$ mK with $\kappa_c/T = 0$ for zero-temperature limit. Since YMGO is a two-dimensional spin system with negligibly weak spin interaction along the *c* axis, the $\kappa_c$ should represent a pure phonon heat transport of this material. This $T^2$ behavior for phonon heat transport is further verified by the high-magnetic-field data of $\kappa_a$. As shown in Figs. 1a and 1b, with applying 14 T (or 10 T) field along the *a* axis (or the *c* axis), the $\kappa_a$ displays a good $T^2$ behavior at $T < 700$ mK accompanied with $\kappa_a/T = 0$ for zero-temperature limit. Since these fields are high enough to polarize all the spins, below the magnon gap, the high-field thermal conductivity should be a purely phononic term without any contribution (carrying heat or scattering phonon) from magnetic excitations. It is therefore reasonable that in high fields there is no residual term of $\kappa_a/T$ at $T \to 0$. Based on above comparisons, it is obvious that in zero field the $\kappa_a$ behaves as $T^2$ at 200 mK $< T <$ 600 mK with a residual thermal conductivity of $\kappa_0/T =$ 0.0058 WK$^{-2}$m$^{-1}$ and thereafter, the slope change leads to a smaller $\kappa_0/T$ = 0.0016 WK$^{-2}$m$^{-1}$ at $T <$ 200 mK. Moreover, the larger $\kappa_a/T$ in high fields indicates that, in zero magnetic field, the phonons are rather strongly scattered by magnetic excitations that are be gapped out in high magnetic fields and suppressed at low temperatures. It firmly suggests the presence of magnetic excitations in the thermal conductivity result at zero magnetic field, and the magnetic excitation certainly does not have a large gap and could be gapless.

It is a bit strange that the phonon thermal conductivity at ultralow temperature has a $T^2$ behavior. Usually, the phonon thermal conductivity of a high-quality insulating crystal at boundary scattering limit should have a $T^3$ temperature dependence. One possible explanation is the surface reflection effect, which could weaken the temperature dependence and gives a $T$-power-law behavior with a power smaller than 3. However, the calculated phonon mean free path (see Supplementary Information) is much smaller than the sample width, at least at several hundreds of millikelvins. Therefore, the surface reflection may not be the origin for the $T^2$ behavior. Here, we would like to leave it as an open question.

The non-zero residual thermal conductivity at zero field immediately implies that YMGO is a QSL with itinerant gapless spin excitations having a long-range algebraic (power-law) temperature dependence. In reality, it is very rare to observe this non-zero $\kappa_0/T$ term in the studied QSL candidates. So far only the organic EtMe$_3$Sb[Pd(dmit)$_2$]$_2$[5,34] and the inorganic 1T-TaS$_2$[41] exhibit a non-zero $\kappa_0/T$ term, both of which are spin-1/2 TAFs. But it is also notable that some recent studies reported a zero $\kappa_0/T$ term in these two materials[42,43]. By following Ref. 5's method, we estimate the mean free path ($l_s$) of the spin excitations in YMGO by calculating

$$\frac{\kappa_0}{T} = \frac{\pi k_B^2}{9\hbar} \frac{l_s}{ad} = \frac{\pi}{9}\left(\frac{k_B}{\hbar}\right)^2 \frac{J}{d}\tau_s \qquad (1)$$

Here, $a$ (~ 3.40 Å) and $d$ (~ 25.1 Å) are the in-plane and out-of-plane lattice constants, respectively. From the observed $\kappa_0/T$ = 0.0058 WK$^{-2}$m$^{-1}$, the $l_s$ is obtained as 78.4 Å, indicating that the excitations are mobile to a distance 23 times as long as the inter-spin distance without being scattered. In comparison, the $\kappa_0/T$ value of YMGO is ~ 30 times smaller than that of EtMe$_3$Sb[Pd(dmit)$_2$]$_2$ (~ 0.19 W/K$^2$m), and ~10 times smaller than that of 1T-TaS$_2$ (~ 0.05 WK$^{-2}$m$^{-1}$), which may be attributed to two possible reasons. First, the spinon velocity is much smaller in YMGO due to the small $J$ value. Second, the $l_s$ is much shorter in YMGO than that in EtMe$_3$Sb[Pd(dmit)$_2$]$_2$ (~1.20 μm)[5], which is related to stronger scattering between phonon and spinon. Note that although 1T-TaS$_2$ has a very large $J$ value[41], its $l_s$ (~50 Å) is also not so long, which may be due to the spin-phonon scattering.

Accordingly, the smaller $\kappa_0/T$ = 0.0016 W/K$^2$m below 200 mK leads to a $l_s$ ~ 21.6 Å, which still covers 6 times of the inter-spin distance. While the exact origin for this slope change (or reduction of the $\kappa_0/T$) is not clear, we suspect it should be related to the effect of disorder. Here, 200 mK is comparable to the AC susceptibility peak position observed at 80 mK. This result strongly suggests that despite the heavily chemical disorder on the Mg/Ga sites, the itinerant spin excitations of YMGO survive when approaching zero temperature.

It is necessary to point out that in an earlier study, a $T^2$ behavior of $\kappa$ measured in the $ab$ plane at $T$ < 300 mK with negative intercept at $T$ = 0 was observed and was interpreted as non-existence of magnetic heat transport[31]. However, if one compares those data with our $\kappa_a$ data, it can be found that those data also exhibit a slope change at ~ 300 mK; the data above 300 mK exhibit a $T^2$ behavior with a non-zero $\kappa_0/T$ = 0.0018 WK$^{-2}$m$^{-1}$. In the present work, first, we measured the thermal conductivity along both the $a$ and $c$ axis. As discussed above, the comparison between the $\kappa_a$ and $\kappa_c$ clearly show the existence of $\kappa_0/T$ term for $\kappa_a$. Second, the phonon mean free path of our sample is much larger than that of Ref. 31 (see Fig. S2 in Supplementary Information). Therefore, it is very clear that our samples display better thermal conductivity, indicating higher sample quality, and should exhibit more intrinsic physical properties of YMGO. Nevertheless, the ultralow-temperature thermal conductivity data from different groups all indicate that at temperatures above 200 or 300 mK, there is itinerant spin excitations associated with the QSL state, while the disorder effect suppresses the transport of spin excitations at lower temperatures.

Figure 2 shows the magnetic-field dependence of $\kappa_a$ at various temperatures and with $B$ // $a$ or $B$ // $c$. First, the $\kappa_a$ is significantly enhanced at high magnetic fields, indicating the existence of magnetic scattering of phonons that can be smeared out in high fields. Second, the $\kappa_a$ exhibits several structures at the low field region. For $B$ // $a$, the $\kappa_a$ isotherm measured at 92 mK shows a minimum at $B_{a1}$ = 0.5 T and another anomaly at $B_{a3}$ ~ 3 T. For $B$ // $c$, two more obvious minima are observed for the 92 mK data at $B_{c1}$ = 0.5 T and $B_{c2}$ = 1.6 T, respectively. These structures under magnetic fields of both directions disappear gradually with increasing temperature.

**Specific heat.** Figure 3 shows the magnetic-field dependence of specific heat ($C_p$) at 300 mK with $B$ // $a$ or $B$ // $c$. For $B$ // $a$, the $C_p$ isotherm shows an obvious slope change at $B_{a3}$ ~ 3.0 T. Its derivative additionally shows another change around $B_{a2} = 1.5$ T, which corresponds to a weaker slope change of $C_p$. For $B // c$, one clear slope change occurs at $B_{c2} = 1.7$ T, which is also highlighted as a peak on its derivative. These features, both shown by the $\kappa$ and $C_p$, pointing to some field-induced crossovers or transitions, or some special field-dependent magnetic scatterings.

**Magnetic torque.** To further investigate the possible transitions, magnetic torque ($\tau$) measurements were performed. Figures 4a and 4b show the calculated $\tau/B$ with applied field along the $a$ or $c$ axis. At 30 mK, the data shows weak hump around 1 ~ 2 T for both directions. In principle, torque magnetometry measures the magnetic anisotropy. Accordingly, the d($\tau/B$)/d$B$ for $B // a$ (Fig. 4c) clearly shows a valley at $B_{a2} = 1.2$ T, a peak at $B_{a3} = 2.7$ T, and another valley at $B_{a4} = 7.0$ T. With increasing temperature, the lower field valley and peak both become weak and disappear with $T \geq 1.5$ K. For $B // c$, the d($\tau/B$)/d$B$ (Fig. 4d) does not show a peak but a flat regime starting around $B_{c2} = 1.7$ T, and a valley at $B_{c3} = 5.0$ T. The $B_{a2}$, $B_{a3}$ and $B_{c2}$ observed here are closely corresponded to the anomaly observed from the $\kappa(B)$ and $C_p(B)$ data. This further confirms the existence of field-induced magnetic crossovers or transitions. It is noticed that the magnetic torque was also measured for YMGO at 350 mK by Steinhardt et al.[44] Its derivative shows broad maxima near 3.5 T for $B // c$ and 5.5 T for $B // a$, which is different from our results obtained at 30 mK while approaching zero temperature.

It is noticed that a recent study reported the DC magnetization measured at 500 mK with $B // ab$ plane[45] or $B // c$. We also measured the 500 mK magnetization with $B // a$ or $B // c$. Our results are the same as the reported ones, as shown in Fig. S4 in Supplementary Information. For $B // a$, the saturation field $B_s$ is around 7.0 T with saturation moment $M_s = 1.45$ $\mu_B$. For $B // c$, $B_s = 5.0$ T and $M_s = 1.8$ $\mu_B$. Therefore, the $B_{a4}$ and $B_{c4}$ represent the saturation fields. It is also noted that the magnetization at $B_{a2} = 1.5$ T, $B_{a3} = 3.0$ T, and $B_{c2} = 1.7$ T is around 1/3, $\sqrt{3}/3$ and 1/2 of the saturation value, respectively, with the assumption that the critical field values are similar between 300 mK and 500 mK. In Ref. 45, they further measured magnetization at 40 mK with $B // c$. Its derivative shows a flat regime around 1.8 T, which is related to 1/2$M_s$. This result is consistent with our observation here. For their data measured at 500 mK with $B // ab$ plane, its derivative shows a flat regime around 3.5 T, which again was ascribed to 1/2$M_s$. This explanation is different from our observation for $B // a$ case, in which both phase boundaries at 1/3$M_s$ and $\sqrt{3}/3M_s$ were observed by the combination of $\kappa(B)$, $C_p(B)$ and $\tau(B)$ data. One reason for this discrepancy could be the fact that the DC magnetization was measured at 500 mK, a relative high temperature. As shown in our data, Fig. 4c, the anomalies on the d($\tau/B$)/d$B$ curve at 30 mK, an ultralow temperature, become weak or flattened pretty quickly with increasing temperature. The $C_p(B)$ data at 200 mK was also reported in Ref. 45. However, its $C_p(B)$ data with $B$ // $ab$ plane shows no distinct feature around 3.0 T, which our $B$ // $a$ data clearly shows. The slope change around 1.5 T, as we observed, also has not been discussed in Ref. 45. Meanwhile, we also noticed that an early study on the DC magnetization measured on powder sample at 500 mK reported two critical fields at 1.6 T and 2.8 T[22], which were suggested as the phase boundaries of a UUD phase. This study supports our observation with $B$ // $a$.

**Phase diagram.** Two magnetic phase diagrams are constructed by using these critical fields, as shown in Figure 5. Since no anomaly was observed on $\tau/B$ around 0.2 ~ 0.5 T, the $B_{a1}$ and $B_{c1}$ are unlikely related to spin state transitions. Accordingly, besides the paramagnetic phase at high temperatures, QSL at low temperature and zero field, and fully polarized phase at high fields, with increasing field, there are three phases for $B$ // $a$ (Fig. 5a) and two phases for $B$ // $c$ (Fig. 5b).

For a spin-1/2 TAF with 120° spin ordering and easy-plane anisotropy, the strong quantum spin fluctuations can stabilize a so-called up-up-down (UUD) phase within a certain magnetic field regime applied in the triangular plane while approaching zero temperature[46-50], which has been observed in Ba$_3$CoSb$_2$O$_9$[51-55], a TAF with effective spin-1/2 Co$^{2+}$ ions. This UUD phase exhibits a magnetization plateau with 1/3 saturated moment ($M_s$). Recently, even in QSL candidates $A$Yb$Ch_2$ ($A$ = Na and Cs, $Ch$ = O, S, Se), one TAF family with effective spin-1/2 Yb$^{3+}$ ions and easy plane anisotropy, the UUD phase has been proposed under applied fields[56-59]. Experimentally, more complex quantum spin state transitions (QSSTs) have been observed for Ba$_3$CoSb$_2$O$_9$[60-66]. For example, with increasing field along the $ab$ plane, its 120° spin structure at zero field is followed by a canted 120° spin structure; the UUD phase; a coplanar 2:1 canted oblique phase with one spin in the 120° spin structure rotated to be parallel with another spin, which gives $\sqrt{3}/3M_s$; and another coplanar phase before entering the fully polarized state.

Since YMGO also has the easy-plane anisotropy[7,8,30], its field-induced anomaly at ultralow temperatures could also be related to these QSSTs. By comparing YMGO's phase diagram with $B$ // $a$ to that of Ba$_3$CoSb$_2$O$_9$, we propose phase I as the canted 120° spin structure, phase II as the UUD phase, and phase III as the oblique phase. The observation of the QSSTs strongly supports the existence of quantum spin fluctuations approaching zero temperature in YMGO, which clearly differentiate it from a conventional spin glass state. It needs to be pointed out is that while the phase boundaries of the UUD phase were observed, the $1/3M_s$ plateau does not exist for YMGO, which could be due to the disorder effect. In another TAF $Rb_{1-x}K_xFe(MoO_4)_2$[67], the disorder introduced by the K-doping also weakens the magnetization plateau feature related to the UUD phase.

It is surprising to see that the phase boundary between phase I and II for $B // c$ case is related to $1/2\,M_s$. As learned from $Ba_3CoSb_2O_9$, while for $B // c$, the 120° spin structure will be followed by an umbrella spin structure and an oblique phase, between which the phase boundary is related to $\sqrt{3}/3M_s$. Meanwhile, Ye and Chubukov calculated the phase diagram of a 2D isotropic triangular Heisenberg antiferromagnet in a magnetic field and predicted a novel up-up-up-down (UUUD) spin sate with a $1/2M_s$ magnetization plateau for $J_2/J_1 > 0.125$ ($J_1$: nearest neighbor exchange interaction, $J_2$: next nearest neighbor exchange interaction)[68]. The reported $J_2/J_1$ values by simulating the spin excitations obtained from INS data[7] and terahertz spectroscopy data[30] are 0.22 and 0.18, respectively. Therefore, we tend to assign the phase II as the UUUD phase, while the nature of Phase I is not clear at this stage. Again, the chemical disorder could be the reason for the disappearance of the $1/2M_s$ plateau.

In the case that the minimum observed at 0.5 T for $\kappa(B)$ belong to some special magnetic scatterings, the scenario of a spinon Fermi surface QSL, that supports gapless magnetic excitations discussed above, may give a possible understanding from the conventional wisdom of Kohn anomaly. Since the charge degrees of freedom in YMGO are frozen out, only Zeeman coupling can be included as the magnetic field is applied. In the weak field regime, the field does not destroy the QSL ground state and the spinon remains to be a valid description of the magnetic excitation[29]. The magnetic fields would modify the spinon Fermi surface, and the Fermi surfaces of spin-up and spin-down spinons expand and shrink with increasing field, respectively. Just like the electron-phonon coupling[69], the spin-lattice coupling may enhance the phonon scatterings with modified spinon Fermi surfaces, resulting in the thermal conductivity modulation. In the high-temperature regime, the QSL breaks down and the structures disappear as in the experiment.

**AC susceptibility.** Finally, we revisited the AC susceptibility ($\chi'$) to check the possibility for a spin glass state in YMGO, although the observation of the residual $\kappa_0/T$ and QSSTs clearly disputes this scenario. Compared to the reported $\chi'$ data performed at zero DC magnetic field and without anisotropic information, our $\chi'$ data was measured with AC field both along the $a$ and $c$ axis, and with applied DC field. Figures 6a and 6b show the $\chi'(T)$ measured with applied DC field along either the $a$ or $c$ axis. With $B = 0$ T, the $\chi'$ shows a peak around 80 mK, which is lower than the reported data exhibiting a peak around 100 mK[35]. With increasing $B$, the intensity of $\chi'$ below 80 mK increases and eventually becomes flat or saturated with $B \geq 0.05$ T. Figures 6c and 6d shows the frequency dependence of the peak position with the AC excitation field either along the $a$ or $c$ axis. For both of them, the peak's position ($T_0$) shift to higher temperatures with increasing frequency. As shown in Figs. 6e and 6f, its frequency dependence can be fit to an Arrhenius law $f = f_0\exp[-E/(k_B T_0)]$, which yields an activation energy $E_a = 3.8(6)$ K and $E_c = 2.5(8)$ K for the excitation field along the $a$ and $c$ axis, respectively. While the frequency-dependent peak of $\chi'$ normally represents a spin glass transition as discussed for YMGO and YbZnGaO$_4$[35], the saturation of the $\chi'$ below this peak position under a small DC field and the anisotropic activation energy both indicate that it should not be treated as a conventional spin glass[70].

Meanwhile, the recent DC susceptibility measurements with $B = 0.01 \sim 0.05$ T for YMGO[27] also revealed a saturation below 100 ~ 200 mK, which has been suggested as a signature for the presence of gapless spin excitations. Since the temperature dependence of the AC and DC susceptibility should behave similarly, the saturation we observed for $\chi'$ could be due to the same origin. In fact, the reported inelastic neutron scattering measurement suggests that at most 16(3)% of the total spectral weight is elastic[7]. Correspondingly, the inelastic contribution (~ 84(3)%) is large compared with the 66% expected for a spin-1/2 glass[71]. Moreover, the DC susceptibility shows no divergence between the zero field cooling and field cooling data, the $\mu$SR relaxation rate shows a plateau below 400 mK[26], and the magnetic entropy has been already released by more than 99% at 80 mK from the specific heat measurement[22]. All these facts again rule out a frozen or glassy state for YMGO. This AC susceptibility peak could originate from the free impurity spins inside or attached to the system.

**Conclusion**

In summary, the observation of a residual $\kappa_0/T$ term and QSSTs strongly supports a QSL state with itinerant spin excitations and quantum spin fluctuations approaching zero temperature in YMGO, although its chemical disorder reduces the mean free path of the excitations and smears out the $1/3M_s$ plateau related to the UUD phase. This survival of QSL state in YMGO is surprising since the Mg/Ga site disorder is supposed to introduce random distribution of exchange interactions and therefore lead to a RS state or even a glassy sate and any field-induced transitions should be expected to smear out completely. Therefore, the chemical disorder in YMGO must play a more complexed role on the exchange interactions. Future studies on the local structure of YMGO to learn how exactly the Mg/Ga disorder affects the Yb-O local environment and the correlated exchange interactions are highly desirable to understand this survival of QSL in YMGO through chemical disorder.

**Methods**

**Sample preparation and characterization.** High-quality YMGO single crystals were grown by using the optical floating-zone technique[8]. By using the X-ray Laue photograph, the crystals were cut precisely along the crystallographic axes for the magnetic and thermal conductivity measurements.

**Thermal conductivity measurements.** Thermal conductivity was measured by using a "one heater, two thermometers" technique in a $^3$He/$^4$He dilution refrigerator at 70 mK – 1 K, equipped with a 14 T superconducting magnet[72,73]. The sample was cut precisely along the crystallographic axes with the longest and the shortest dimensions are along the *a* or the *c* axis. The magnetic fields were applied along either the *a* or *c* axis. Gold paint was used to make four contacts on each sample. The $RuO_2$ chip resistors were used as heaters and thermometers and are connected to the gold contacts by using gold wires and silver epoxy. The $RuO_2$ thermometers were pre-calibrated by using a $RuO_2$ reference sensor (Lakeshore Cryotronics) mounted at the mixture chamber (the superconducting magnet was equipped with a cancellation coil at the height of mixture chamber).

**AC susceptibility measurements.** The ac susceptibility was measured using the conventional mutual inductance technique (with a combination of ac current source and a lockin amplifier) at SCM1 dilution fridge magnet of the National High Magnetic Field Laboratory, Tallahassee. The typical AC field strength is 1.1-1.6 Oe).

**Torque measurements.** Torque magnetometry was performed at the University of Michigan using a self-built capacitive cantilever setup mounted inside a dilution refrigerator[74-76]. The sample was mounted on a thin BeCu cantilever. To infer the magnetic torque $\tau$, the cantilever deflection is measured by tracking the change of the capacitance between the cantilever and a gold thin film underneath. The report magnetization is the effective torque magnetization $M = \tau/B$.

**Specific heat measurements.** The specific heat was measured with a Quantum Design physical property measurements system (PPMS), equipped with a dilution refrigerator insert.

**DC magnetization measurements.** The DC magnetization curves was measured with a Quantum Design SQUID-VSM, equipped with a $^3$He refrigerator insert.

**Data availability**

The data that support the findings of this study are available from the corresponding authors upon reasonable request.

**Acknowledgements**

This work was supported by the National Natural Science Foundation of China (Grants No. U1832209, No. 11874336, No. 12025408, and No. 11921004), the National Basic Research Program of China (Grants No. 2016YFA0301001, No. 2018YFGH000095, No. 2016YFA0300500, and No. 2018YFA0305700), the Innovative Program of Hefei Science Center CAS (Grant No. 2019HSC-CIP001), and the Research Grants Council of Hong Kong with General Research Fund (Grant No.17303819). Y.S. and J.Z. were supported by the Innovation Program of Shanghai Municipal Education Commission (Grant No. 2017-01-07-00-07-E00018) and the National Key RD Program of the MOST of China (Grant No. 2016YFA0300203). Z.D., Q.H., and H.Z. thank the support from National Science Foundation through award DMR-2003117. T.A.,

L.C., and L.L. (the torque magnetometry work at Michigan) thank the support from the U.S. Department of Energy (DOE) under Award No. DE-SC0020184. A portion of this work was performed at the National High Magnetic Field Laboratory, which is supported by the National Science Foundation Cooperative Agreement No. DMR-1644779 and the State of Florida.

**Author contributions**

X.R., G.H., W.J.C., N.L., and X.F.S. performed thermal conductivity measurements and analyzed the data with help from X.Z., G.C., and H.D.Z. Q.H., Z.D., E.S.C., and H.D.Z. performed the low-temperature AC susceptibility measurements. T.A., L.C. and L.L. performed the torque measurements. X.Y.Y. measured the specific heat. N.N.W. and J.G.C. measured the DC magnetization. Y.H.G., Y.S., and J.Z. made the samples. X.F.S., H.D.Z., and G.C. wrote the paper with input from all other co-authors.

**Competing interests**

The authors declare no competing interests.

**Additional Information**

**Correspondence and requests for materials** should be addressed to X.F.S., H.D.Z. or C.G.

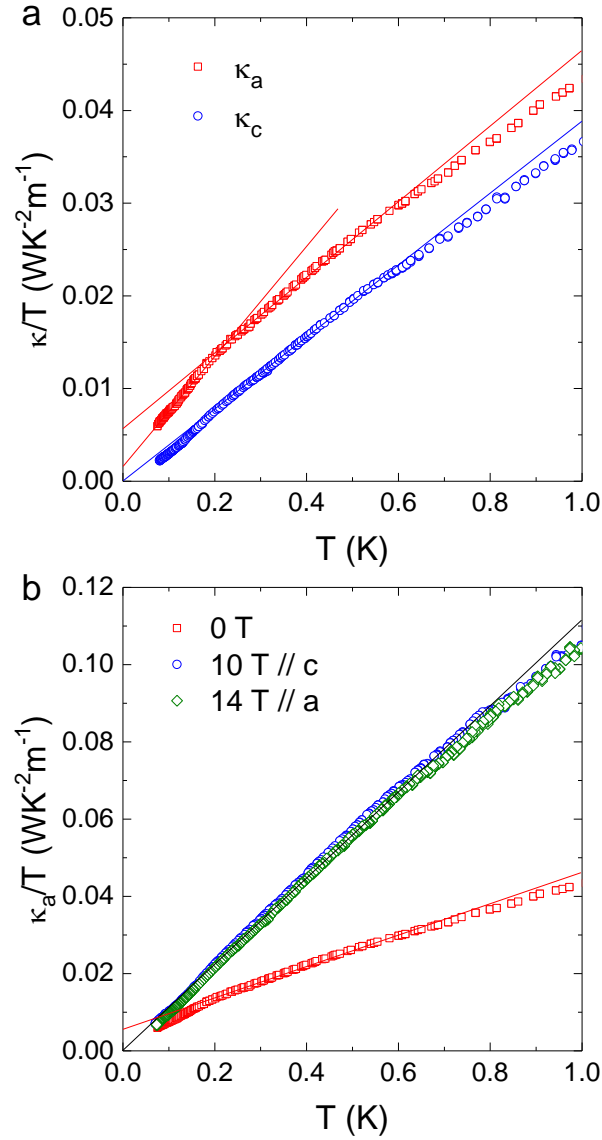

**Figure 1 Ultralow-temperature thermal conductivity of YbMgGaO$_4$ single crystals. a,** Data measured along the *a* axis ($\kappa_a$) and the *c* axis ($\kappa_c$), plotted as $\kappa/T$ vs $T^2$. The solid lines are some linear fitting results. The $\kappa_c$ display a $T^2$ behavior at $T < 600$ mK with zero intercept at $T = 0$. The $\kappa_a$ behave as $T^2$ at 200 mK $< T <$ 600 mK with a residual thermal conductivity of $\kappa_0/T = 0.0058$ WK$^{-2}$m$^{-1}$ and a smaller $\kappa_0/T = 0.0016$ WK$^{-2}$m$^{-1}$ at $T <$ 200 mK. **b,** The *a*-axis thermal conductivity in zero field and in 10 T (14 T) magnetic field applied along the *a* (the *c*) axis. The high-field data The $\kappa_c$ display a $T^2$ behavior at $T < 700$ mK with $\kappa_0/T = 0$.

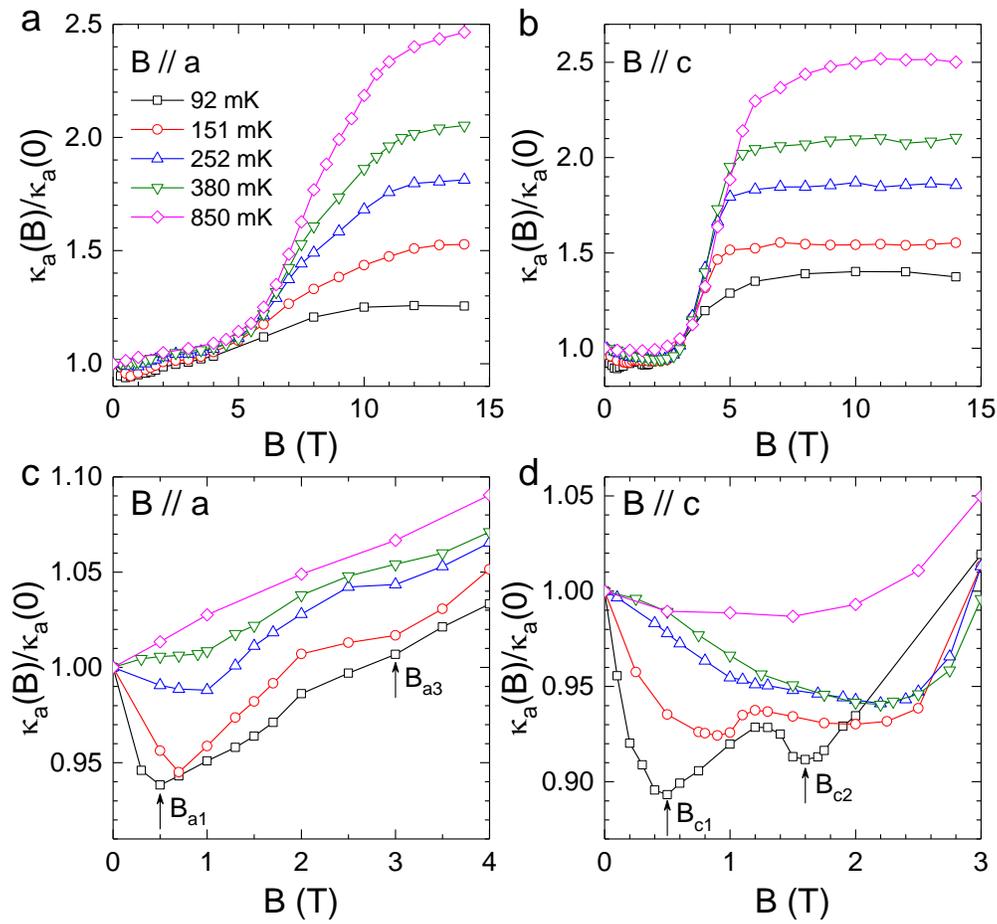

**Figure 2 Magnetic-field dependence of the *a*-axis thermal conductivity of YbMgGaO$_4$. a,b,** Magnetic field dependence of $\kappa_a$ at various temperatures with field applied along either the *a* axis or the *c* axis. **c,d,** A magnified view of $\kappa_a(B)$ data at low fields. At 92 mK and with *B* // *a*, there is a minimum at 0.5 T and a slope change at 3 T, indicated by arrows. While at 92 mK and with *B* // *c*, there are two minima at 0.5 and 1.6 T. With increasing temperature, these anomalous minima become weaker and disappear above 850 mK.

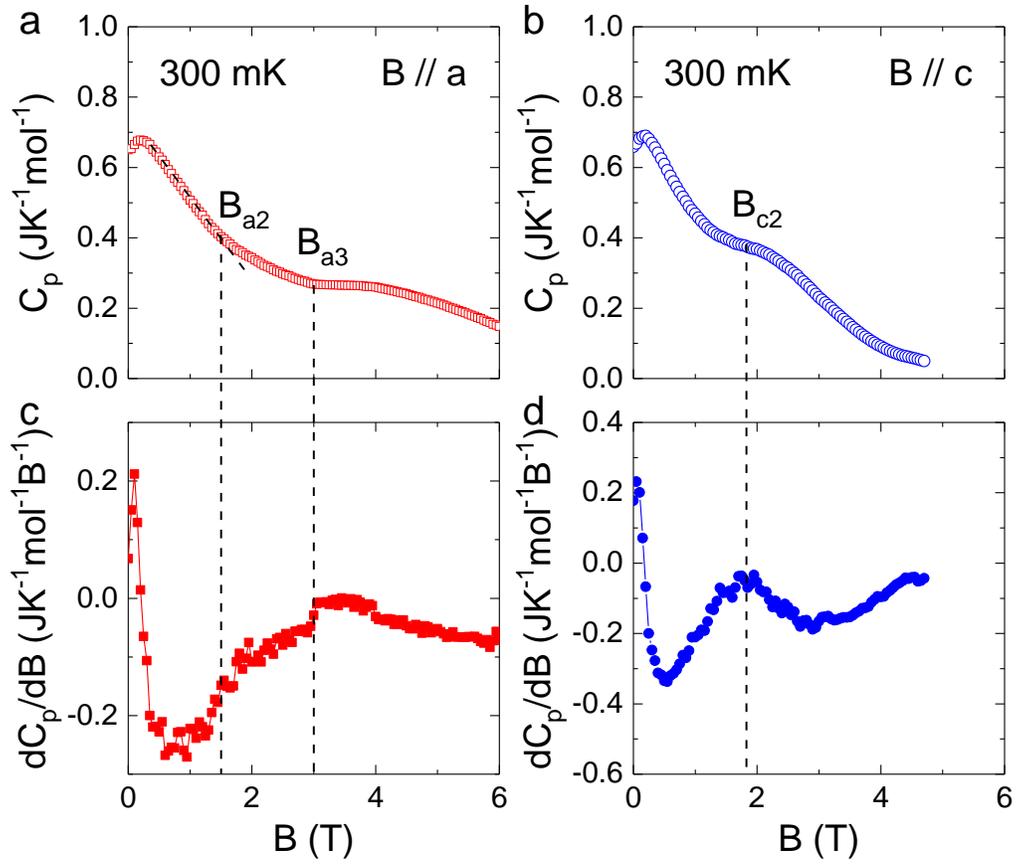

**Figure 3 Low-temperature specific heat of YbMgGaO$_4$. a,b,** The magnetic-field dependence of specific heat, $C_p$, measured at 300 mK and with $B // a$ or $B // c$. **c,d,** The derivative ($dC_p/dB$) for $B // a$ or $B // c$. The dashed lines are the guide for eyes and indicate the anomalies associated with some transition fields $B_{a2}$, $B_{a3}$, and $B_{c2}$.

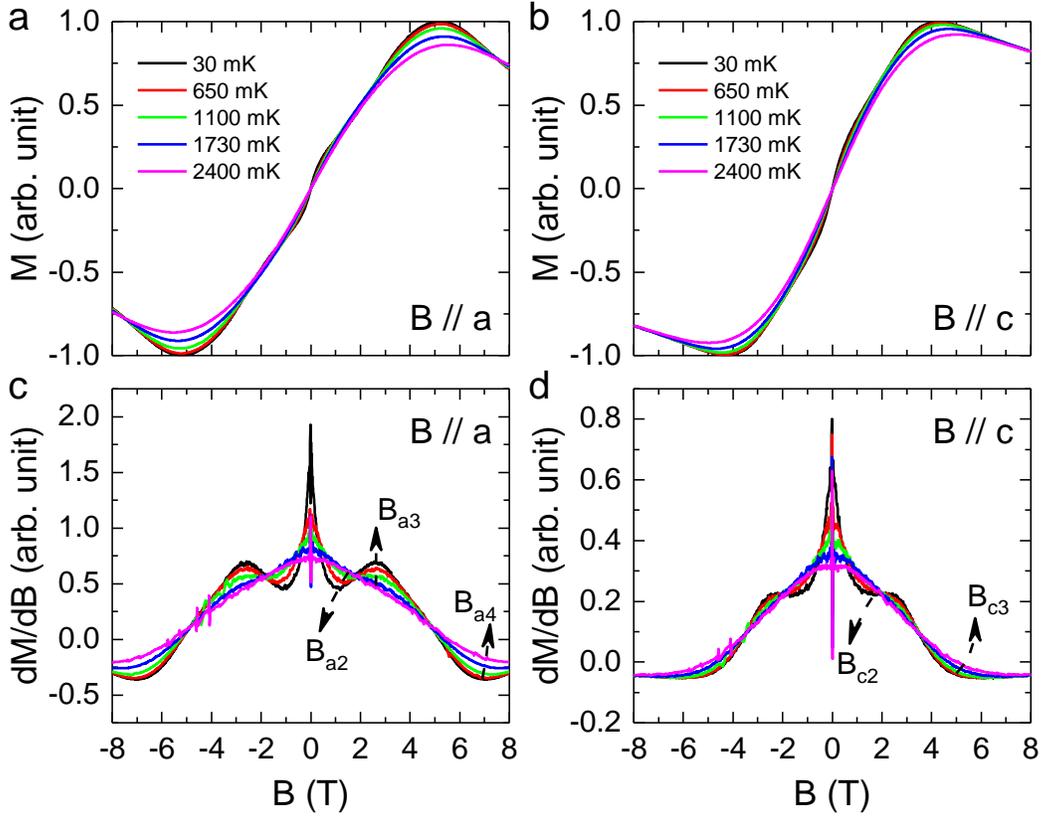

**Figure 4 Ultralow-temperature magnetization of YbMgGaO$_4$. a,b,** The calculated magnetization ($M$) from the measured magnetic torque as torque/$B$ with field applied along either the $a$ axis or the $c$ axis (See Supplementary Figure 1). **c,d,** The derivative (d$M$/d$B$) for $B \mathbin{/\mkern-2mu/} a$ or $c$ axis. At 30 mK, the derivative shows two anomalies at $B_{a2}$ and $B_{a3}$ for $B \mathbin{/\mkern-2mu/} a$, which are related to $1/3 M_s$ and $\sqrt{3}/3 M_s$, respectively. Meanwhile, only one anomaly at $B_{c2}$ was observed from the derivative for $B \mathbin{/\mkern-2mu/} c$, which is related to $1/2 M_s$.

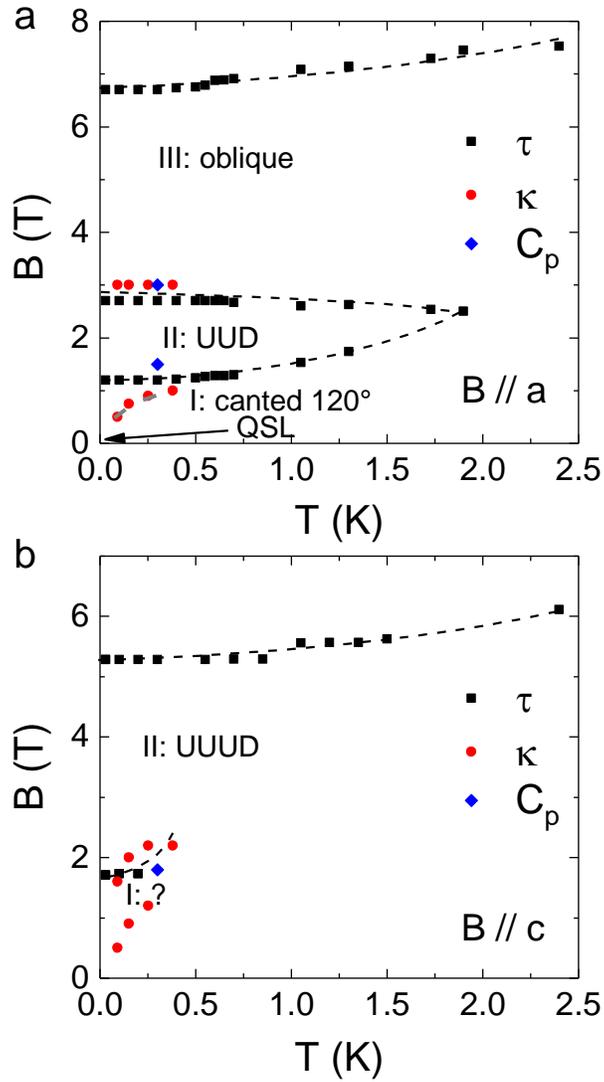

**Figure 5 Magnetic phase diagrams of YbMgGaO$_4$. a,** for $B // a$. **b,** for $B // c$. The data points are obtained from the magnetic torque ($\tau$) and field dependence of thermal conductivity ($\kappa$) measurements. The dashed lines are phase boundaries. For $B // a$, there are three phases (I: thee canted 120° spin structure, II: the UUD phase, and III: the oblique phase) in the low-temperature and low-field region. Whereas, there are two phases (I: unknown phase and II: UUUD phase) at low temperature and low field for $B // c$. The dashed lines are the guide to the eyes.

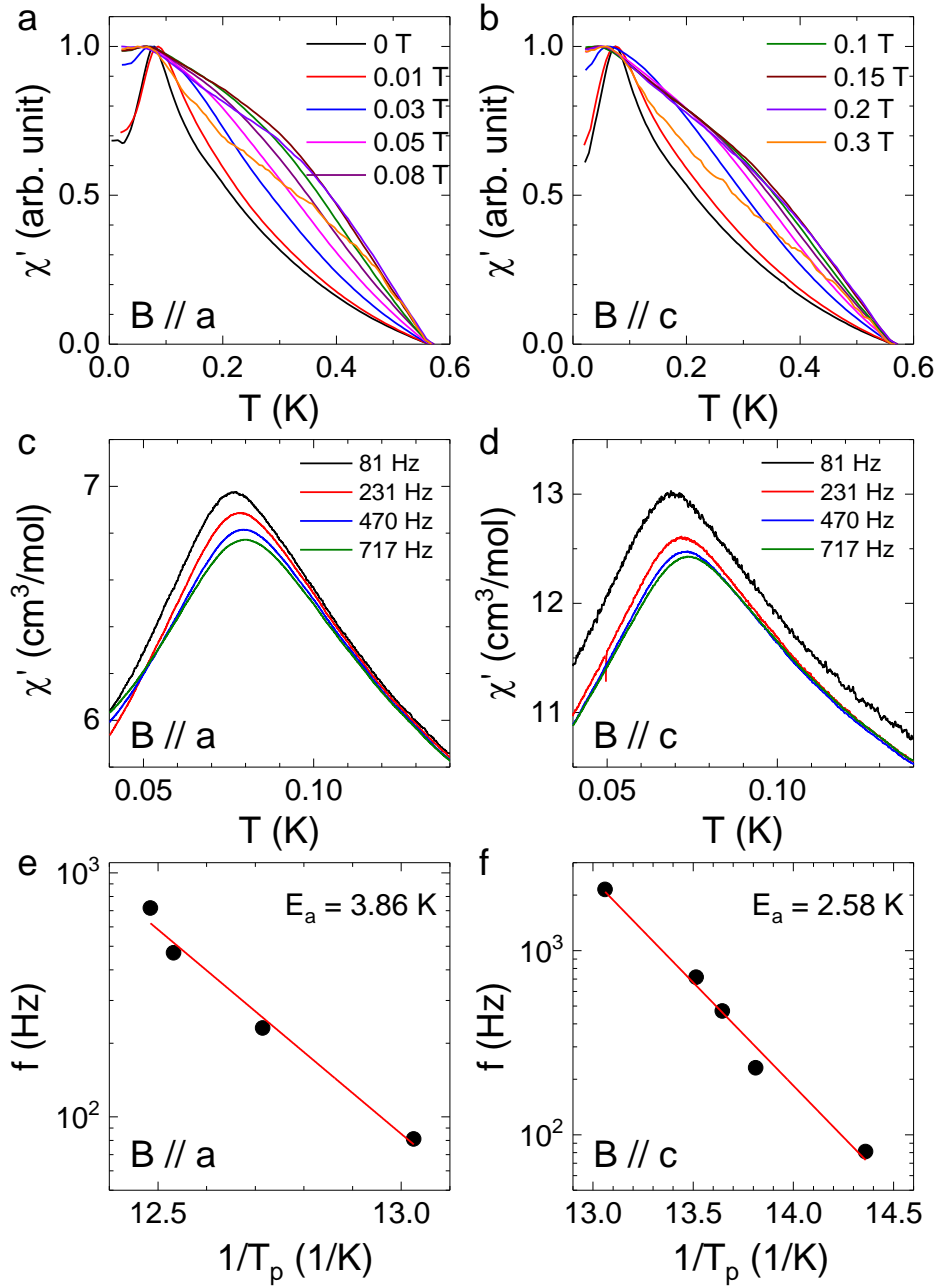

**Figure 6 AC susceptibility of YbMgGaO$_4$. a,b,** Temperature dependence of the real part of the AC susceptibility, $\chi'$, measured with different DC magnetic fields along either the *a* axis or the *c* axis. Here, the maximum value of each data was scaled to 1 to clearly show the DC magnetic field effects on AC susceptibility. **c,d,** Frequency dependence of the $\chi'$ peak. Here, the absolute value of the AC susceptibility was obtained by rescaling it to the DC susceptibility (see Supplementary Information). **e,f,** Arrhenius law fit of the $\chi'$ peak position, $T_0$. The applied AC excitation field (~ 1 Oe) is along either the *a* axis for **(c,e)** or the *c* axis for **(d,f)**.

# Supplementary Information for
## *"Survival of itinerant excitations and quantum spin state transitions in YbMgGaO₄ with chemical disorder"*

X. Rao[1,10], G. Hussain[1,10], Q. Huang[2,10], W. J. Chu[1], N. Li[1], X. Zhao[3], Z. Dun[2], E. S. Choi[4], T. Asaba[5], L. Chen[5], L. Li[5], X. Y. Yue[6], N. N. Wang[7], J.-G. Cheng[7], Y. H. Gao[8], Y. Shen[8], J. Zhao[8], G. Chen[9]⋆, H. D. Zhou[2]⋆, and X. F. Sun[1,6]⋆

[1]Hefei National Laboratory for Physical Sciences at Microscale, Department of Physics, and Key Laboratory of Strongly-Coupled Quantum Matter Physics (CAS), University of Science and Technology of China, Hefei, Anhui 230026, People's Republic of China

[2]Department of Physics and Astronomy, University of Tennessee, Knoxville, Tennessee 37996-1200, USA

[3]School of Physical Sciences, University of Science and Technology of China, Hefei, Anhui 230026, People's Republic of China

[4]National High Magnetic Field Laboratory, Florida State University, Tallahassee, FL 32310-3706, USA

[5]Department of Physics, University of Michigan, Ann Arbor, Michigan 48109, USA

[6]Institute of Physical Science and Information Technology, Anhui University, Hefei, Anhui 230601, People's Republic of China

[7]Beijing National Laboratory for Condensed Matter Physics and Institute of Physics, Chinese Academy of Sciences, Beijing 100190, People's Republic of China

[8]State Key Laboratory of Surface Physics and Department of Physics, Fudan University, Shanghai 200433, People's Republic of China

[9]Department of Physics and HKU-UCAS Joint Institute for Theoretical and Computational Physics at Hong Kong, The University of Hong Kong, Hong Kong, China

[10]These authors contributed equally: X. Rao, G. Hussain, Q. Huang

⋆email: gangchen.physics@gmail.com; hzhou10@utk.edu; xfsun@ustc.edu.cn

*Torque data*:

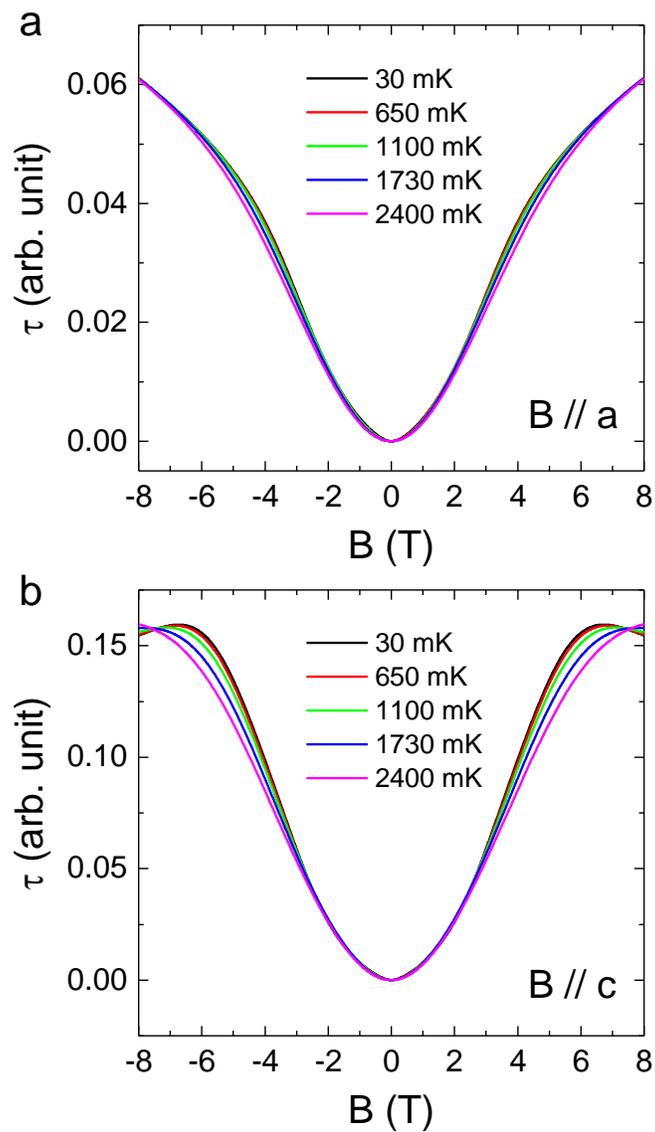

**Supplementary Figure S1** Field dependence of the magnetic torque of YbMgGaO$_4$ at different temperatures. **a,** for *B // a*. **b,** for *B // c*.

*Phonon mean free path:*

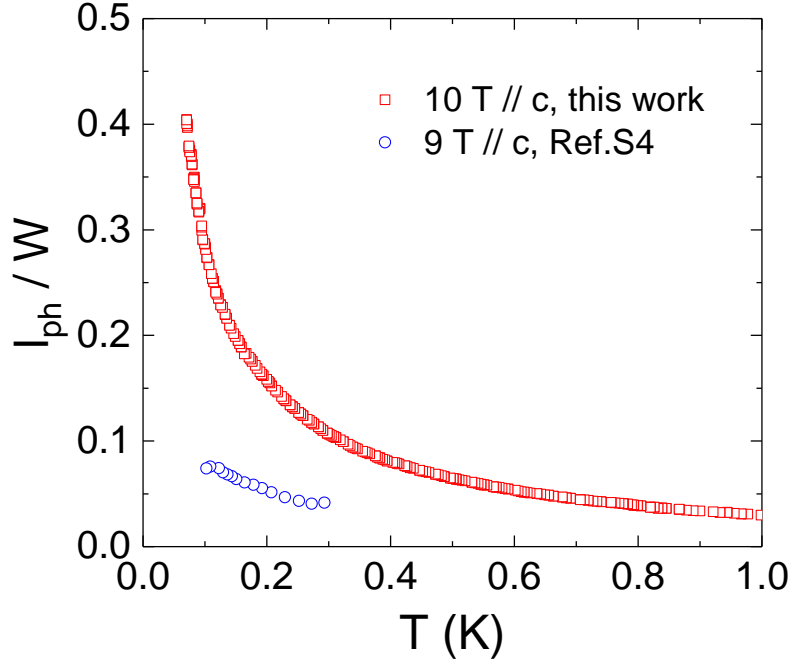

**Supplementary Figure S2** Temperature dependence of the phonon mean free path $l_{ph}$ divided by the averaged sample width $W$, calculated from our $\kappa_a$ sample with 10 T // $c$ and the sample in Ref. S4 with 9 T // $c$.

It is notable that since all the thermal conductivity data (including both $\kappa_a$ and $\kappa_c$, zero field and high field) display a temperature dependence close to $T^2$. Usually, the phonon thermal conductivity of a high-quality insulating crystal at boundary scattering limit should have a $T^3$ temperature dependence. One possible reason is related to the surface reflection effects, which could result in phonon mean free path long than the sample size and a $T$-power-law behavior with a power smaller than 3. Therefore, it is useful to calculate the phonon mean free path from our thermal conductivity data. First, we analyzed the specific heat data of LuMgGaO$_4$, which are purely phononic and can be a reference of the phonon specific heat of YbMgGaO$_4$. We got the raw data (at 0.26–30 K) from Ref. S1 and fitted the data by using the low-temperature expansion of the Debye function, $C = \beta T^3 + \beta_5 T^5 + \beta_7 T^7$ (see Refs. S2 and S3). The fitting parameters are $\beta = 5.19 \times 10^{-4}$ J/K$^4$mol, $\beta_5 = -3.24 \times 10^{-7}$ J/K$^6$mol, and $\beta_7 = 6.60 \times 10^{-11}$ J/K$^8$mol. The phononic thermal conductivity can be expressed by the kinetic formula $\kappa_{ph} = 1/3 C v_{ph} l_{ph}$, where $C = \beta T^3$ is phonon specific heat at low temperatures, $v_{ph}$ is the average velocity, and $l_{ph}$ is the mean free path of phonon. Here $\beta = 5.19\times10^{-4}$ J/K$^4$mol is obtained from the above specific-heat data and $v_{ph} = 2070$ m/s can be calculated from $\beta$ (see Ref. S3). Then, we can calculate $l_{ph}$ from the $\kappa_a(T)$ data at 10 T field (// $c$) and compare it with the averaged sample width $W = 2(A/\pi)^{1/2} = 0.375$ mm (for the $\kappa_a$ sample), where $A$ is the area of cross section. For comparison, we also analyzed the data in Ref. S4 (in-plane thermal conductivity with 9 T // $c$) and calculated the mean free path of phonons. In Figure S2, we plot the temperature dependence of $l_{ph}/W$ of our $\kappa_a$ sample with 10 T // $c$ and the sample in Ref. S4 with 9 T // $c$. Apparently, our samples display better thermal conductivity, indicating higher sample quality, and should exhibit more intrinsic physical properties of YMGO.

Although the phonon mean free path keeps increasing with decreasing temperature, it is still smaller than the averaged sample width at the lowest temperatures. This might be related to some uncertainties in the above fitting and calculations, such as the determination of the $\beta$ coefficient, the slight difference in phonon specific heat between YbMgGaO$_4$ and LuMgGaO$_4$, etc. However, it is more likely that the phonon mean free path is much smaller than the sample width, at least at several hundreds of millikelvins. Therefore, the surface reflection may not be the origin for the $T^2$ behavior of thermal conductivity.

Except for the surface reflection effect, it is very hard to understand a $T^2$ behavior of phonon thermal conductivity at such low temperatures. Usually at subkelvin temperatures, phonons are free from the microscopic scattering by lattice imperfections, like point defects and dislocations, and $\kappa_{ph}$ displays a temperature dependence close to $T^3$. One may suspect that it is due to the magnetic scattering effect. This scattering is indeed playing a role considering the rather strong magnetic-field dependence of $\kappa$. However, high magnetic field is believed to suppress the magnetic excitations and smear out the magnetic scattering effect. Thus, it is still a mystery why the high field data, which is purely phononic, display a $T^2$ behavior. We would like to leave it as an open question.

Nevertheless, since all the data (including both $\kappa_a$ and $\kappa_c$, zero field and high field) display a temperature dependence close to $T^2$, it is reasonable to analyze the low temperature data by using the $\kappa/T$ vs $T$ and to get the intercept values at $T = 0$.

*Magnetic susceptibility scaling*:

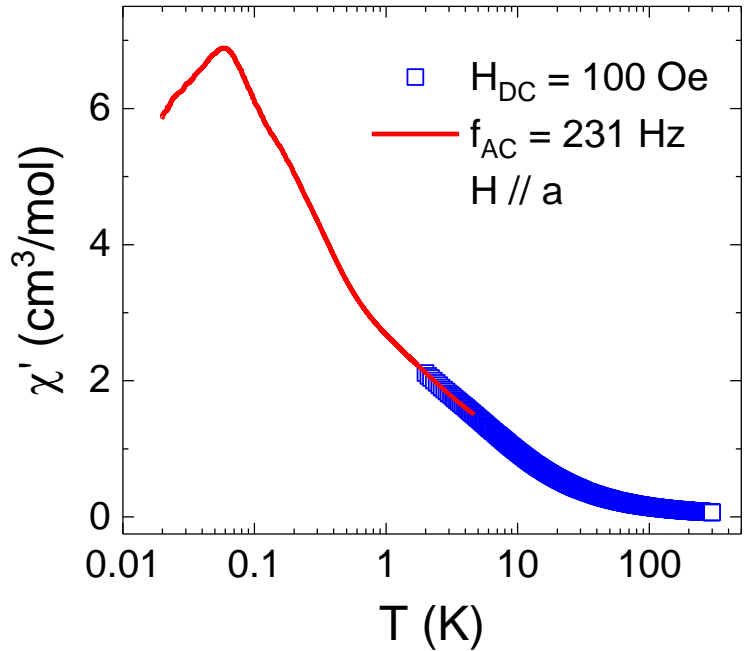

**Supplementary Figure S3** The magnetic susceptibility for YbMgGaO$_4$ shown with DC data and scaled AC data.

The temperature dependence of the AC susceptibility measured with a small AC field and low frequency should reflect the intrinsic susceptibility behavior of a system, or has the similar temperature trend of the DC susceptibility measured on the same system. Therefore, a AC field of 1 Oe with frequency 231 Hz was used to measure the magnetic susceptibility down to 30 mK. This data was easily matched to the high temperature DC susceptibility data measured with a DC a field at 100 Oe and taken down to 1.8 K with a simple scaling factor. Unit of the DC susceptibility, cm$^3$/mol, is used for scaled AC data to maintain continuity. The data in Figures 5c and 5d of the main text with absolute value was obtained by this scaling.

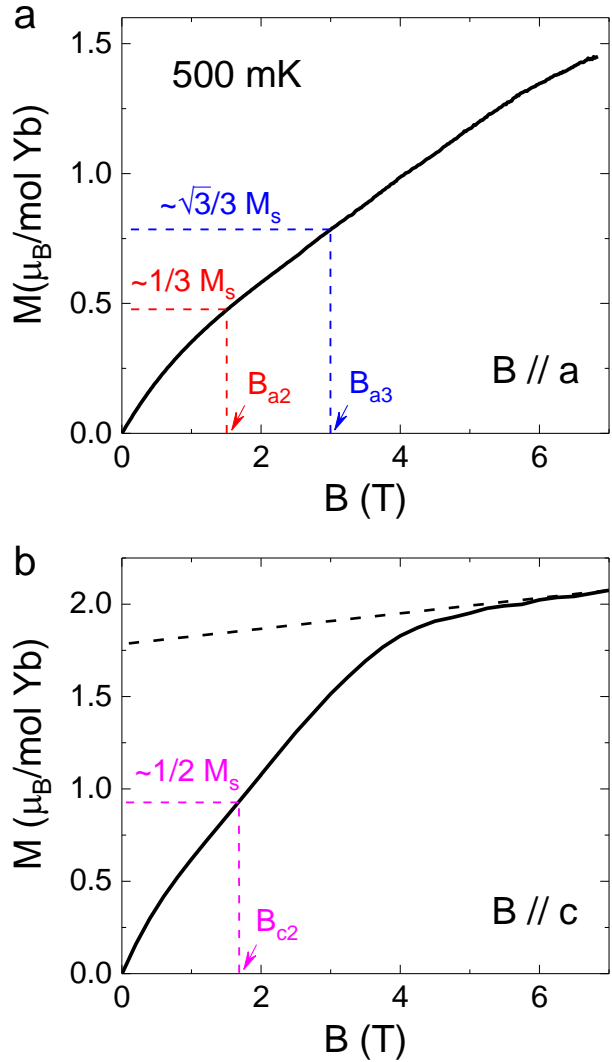

**Figure S4 DC magnetization of YbMgGaO$_4$ at 500 mK. a,** for $B \parallel a$. **b,** for $B \parallel c$. The magnetization at $B_{a2}$ = 1.5 T, $B_{a3}$ = 3.0 T, and $B_{c2}$ = 1.7 T is around 1/3, $\sqrt{3}/3$ and 1/2 of the saturation value, respectively. The thick dashed line in **b** indicates the Van Vleck paramagnetic background.